\documentclass[showpacs,preprintnumbers,pre]{revtex4}
\usepackage{graphicx}
\usepackage{dcolumn}
\usepackage{latexsym,amsmath,amssymb,graphics,color}
\usepackage{bm}
\usepackage{epsf}
\usepackage[centerlast,footnotesize,sf]{caption}
\usepackage{multirow}
\usepackage{subfigure}

\begin{document}

\title{Comparison between different oxygen adsorption mechanisms for the catalytic oxidation of CO on a surface}

\author{C.~Ojeda
}
\author{
G.~M.~Buend{\'{\i}}a}\email{buendia@usb.ve}

\affiliation{
Department of Physics, Universidad Sim{\'o}n Bol{\'{\i}}var, Caracas 1080, Venezuela}

\date{\today}
	
\maketitle

We study by kinetic Monte Carlo simulations the dynamic behavior of a Ziff-Gulari-Barshad (ZGB) model for the catalytic oxidation of CO on a surface. It is well known that the ZGB model presents a continuous transition between an oxygen poisoned state and a reactive state that it is not observed in nature. Based on some experimental results that indicate that the oxygen atoms move away from each other upon dissociation at the surface, we modify the standard ZGB model by changing the entrance mechanism of the oxygen molecule. We study three different ways in which the oxygen atoms can be adsorbed at the surface such that the nonphysical continuous phase transition disappears. We calculate the phase diagram for the three cases and study the effects of including a CO desorption mechanism.

\section{Introduction}
Catalytic reactions on surfaces have attracted a great deal of interest. Besides their enormous technological and industrial applications, they present a rich and complex behavior that makes them an ideal laboratory to study non-equilibrium phenomena ~\cite{catalytic,imbhil95,marr99}.

The Ziff-Gulari and Barshad (ZGB) model ~\cite{ziff86}, is a simple but fruitful system to describe the catalytic oxidation of carbon monoxide (CO) on a surface. In this model the catalytic surface is represented by a square lattice whose sites can be empty or filled with oxygen (O) or CO.  When CO and O are located on nearest-neighbors sites on the lattice they react and liberate CO$_2$, leaving behind two empty sites on the surface, i.e., a Langmuir-Hinshelwood  mechanism. The kinetic of the reaction is described by a single parameter, $y$, the probability that the next molecule arriving to the surface is a CO. The model presents two phase transitions, a continuous one at $y_1$ between a passivated phase where the surface is completely filled with oxygen and a reactive phase, and a discontinuous one at $y_2$ between the reactive phase and a passivated phase with the surface filled with CO. However, experimental results indicate that above a critical temperature, the discontinuous transition disappears and the CO$_2$ production decreases continuously ~\cite{ehsa89b}. This behavior can be reproduced by modifying the ZGB model to include a CO desorption rate, $k$, that mimics the effect of the temperature ~\cite{kauk89, bros92, mach05a, buen06} (it has been shown experimentally that the desorption rate of the oxygens is negligible ~\cite{ehsa89}). Experiments also indicate that in real systems there is not a passivated phase where the surface is poisoned with O, the reaction rate increases as soon as $y$ departs from zero ~\cite{ehsa89,kris92}. More realistic treatments of the catalytic oxidation are oriented toward modifying the ZGB model to eliminate the nonphysical transition at $y_1$. Several authors have shown that this transition can be eliminated by adding to the Langmuir-Hinshelwood an Eley-Rideal step that allows a reaction between CO molecules in the gas phase and adsorbed O atoms on the surface ~\cite{meak90, tamb94, buen09}. However, there are no experimental data indicating that such a reaction between free CO and adsorbed oxygen occurs. Inspired by experiments based on scanning tunneling microscopy that point toward a "hot atom" mechanism according to which the two oxygen atoms are propelled apart upon dissociation ~\cite{wint96}, we show that the nonphysical transition at $y_1$ can be eliminated by making simple changes to the entrance mechanism of the oxygen atoms in the ZGB model. Similar mechanisms have been suggested by other authors ~\cite{khan04, khal06, alba94}. In the ZGB model upon adsorption the oxygen atoms are located one lattice unit apart. In this work we compare three different entrance mechanisms for the oxygen atoms. In the first case, upon dissociation the oxygen atoms are separated by $\sqrt 2$ lattice units, in the second they are located at opposite sites of an empty site, separated by $2$ lattice units, and in the third case their locations are the same as in the second but the center site does not need to be empty. All these mechanisms have the advantage of incorporating the hot atom effect without changing the nature of the model. In this work we perform Monte Carlo simulations to calculate the phase diagrams of the different models. Then, in order to include temperature effects, we add a desorption probability for the CO molecules and determine how the phase diagrams are affected.

\section{The models and the numerical simulation}

The models are simulated on a square lattice of
linear size $L$ that represents the catalytic surface. A Monte Carlo
simulation generates a sequence of trials:  A CO or O$_2$ molecule is selected
with probability $y$ and $1-y$ respectively ~\cite{ziff86}.
These probabilities are proportional to the partial pressures of the gases.
The algorithm works in the following way. A site $i$ is selected
at random. If $i$ is empty and CO adsorption is selected (probability $y$), a CO 
molecule is adsorbed. If one of the nearest-neighbors (nn) of $i$ is filled with an O, there is a reaction between the CO and the O: a CO$_2$ molecule is liberated leaving two empty nn sites at the surface. The O$_2$ adsorption requires two empty sites on the lattice to accommodate the two dissociated atoms. There are several ways to choose these sites. The main purpose of this work is to compare these different adsorption mechanisms. In the standard ZGB model, if the site $i$ is empty and 0$_2$ adsorption is performed (probability $1-y$), a nearest neighbor of $i$ is selected at random, if it is empty the adsorption takes place: one O is adsorbed at $i$ and the second at the chosen nn site. Another option is to adsorb the two oxygens at next-nearest neighbors (nnn) sites: the adsorption proceeds only if one of the nnn of the empty site $i$ is also empty, this case is going to be labeled the ZGB-n model. A different mechanism that we are going to label the Ha1 model, consists in performing the adsorption only if the selected site $i$ is empty and one of the vertical or horizontal nn pairs of the site $i$ are empty. If this is the case an O is located in each site of the pair (leaving the site $i$ empty). In a slightly different version, that will be labeled Ha2, the site $i$ does not need to be empty for the adsorption to proceed in one of its nn pairs. For all the models, after an O$_2$ adsorption has proceed, all the nn of the adsorbed oxygens are examined. If any of them is filled with an CO a reaction proceeds, and a CO$_2$ molecule is liberated. The different adsorption mechanisms for O$_2$ are sketched in Fig.~\ref{sk1} and Fig.~\ref{sk2}. By locating the adsorbed oxygens further apart than in the original ZGB model, in a certain way all the new mechanisms take into account the hot-atom effect.
A schematic representation of the ZGB model is the following,

\begin{eqnarray}
\text{CO(g)} + \text{S} & \rightarrow & \text{CO(a)}
\nonumber \\
\text{O}_2 + 2\text{S} & \rightarrow & 2\text{O(a)}
\nonumber \\
\text{CO(a)} + \text{O(a)} & \rightarrow & \text{CO}_2 \text{(g)} + 2\text{S}
\nonumber 
\end{eqnarray}

Here $S$ represents an empty site on the surface, $g$ means gas phase and $a$ means adsorbed.
For our simulations we assume periodic boundary conditions. The
time unit is one Monte Carlo Step per Site, MCSS, in which each
site is visited once, on average. Averages are taken over $3\times10^4$ MCSS for each set of parameters after $2\times10^4$ warming steps.

\section{Results}
Starting from an empty lattice of size 100$\times$100, we wait until the system reaches a steady state at a constant partial pressure $y$. We calculate the
fraction of sites occupied by CO molecules, the CO coverage 
($\theta_{\rm CO}$) and the O coverage ($\theta_{\rm O}$). $R_{\rm{CO}_2}$ is the rate of production of CO$_2$. 
In Fig.~\ref{fig1}(a) we show the CO coverage, in Fig.~\ref{fig1}(b) the O coverage, and in Fig.~\ref{fig2} the production rate, for the models analyzed. The first thing to notice is that for the ZGB-n, Ha1, and Ha2 models, the continuous transition at $y_1$ that presents the ZGB model is eliminated: the CO$_2$ production starts as soon as $y$ departs from zero (see inset of Fig.~\ref{fig2}). The fact that for $y>0$ the surface can not be completely filled with O is obvious for the Ha1 model: the entrance of oxygens requires the existence of empty sites, empty sites that can be filled with a CO and then react and create more empty sites. For the Ha2 and ZGB-n models the reason is more subtle: each CO-O reaction produces two empty nn sites, that contrary to what happens in the ZGB model, can not be filled by oxygens. The change in the entrance mechanisms for the oxygens also affects the location of the discontinuous transition at $y_2$ that now has been shifted toward lower values of $y$. This is an expected result, since it is now easier for the surface to be filled by CO. It is not surprising  that this effect is particularly strong for the Ha1 model. Notice that, comparing with the ZGB model, for the ZGB-n and the Ha2 models there is a slight increase of the maximum value of the CO$_2$ production that occurs very close to the $y_2$ transition, while for the Ha1 model the production increases very fast as soon as $y$ departs from zero but the maximum value reached is lower than the others. The reactive windows of the ZGB-n and the Ha2 models are larger than the one for the ZGB model.

Next we include the effect of CO desorption by adding a reaction of the form,

\begin{equation}
\text{CO(a)}  \rightarrow  \text{CO(g)}  + \text{S}
\nonumber
\end{equation}

Now we perform a CO or O$_2$ adsorption with probability 1-$k$ and a CO desorption with probability $k$. As we already mentioned, this reaction mimics the effect of the temperature in real systems. In Fig.~\ref{fig3} we present the CO coverages for $k=0.02$ (a) and $k=0.06$ (b). In Fig.~\ref{fig4} we present the O coverages for $k=0.02$ (a) and $k=0.06$ (b). In Fig.~\ref{fig5} we present the CO$_2$ production rates for $k=0.02$ (a) and $k=0.06$ (b). One effect of including the desorption term is that the $y_2$ transition is slightly shifted toward higher values of $y$. This is expected since the CO-desorption reaction makes it more difficult for the surface to be poisoned with CO. Also as can be seen in Fig.~\ref{fig6}(a) and (b), the production rate increases with the value of $k$. Notice that the transition at $y_2$ seems to become smoother as $k$ increases, this is consistent with experimental results and previous simulations that indicate that, for the ZGB model with CO desorption, there is a limiting value of $k$ at which the first-order transition line terminates ~\cite{ehsa89b,tome93,mach05}. Our results suggest that, if there is such a critical $k$ for the new models, it is higher than the one for the ZGB.

\section{Conclusions}
\begin{figure}[ht]
\includegraphics[clip,angle=0,width=.5 \textwidth]{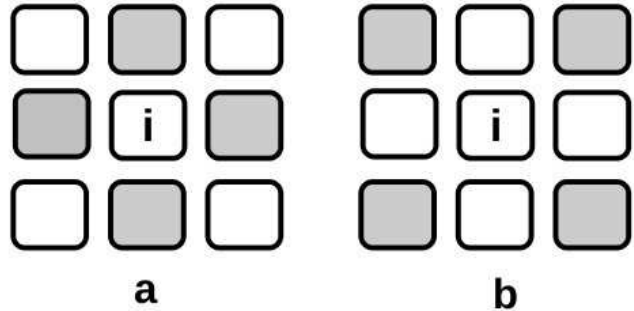}
\caption{ O$_2$ adsorption mechanisms. One O is adsorbed at the initially empty site $i$. a) ZGB model: the second O is adsorbed at a nn of $i$. b) ZGB-n model: the second O is adsorbed at a nnn of $i$. }
\label{sk1}
\end{figure}

\begin{figure}[ht]
\includegraphics[width=14pc]{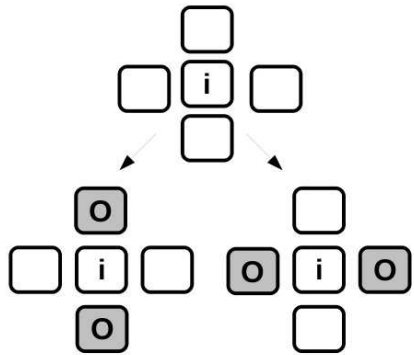}
\caption{ O$_2$ adsorption mechanisms for the Ha1 and Ha2 models. The two O atoms are adsorbed in the vertical or horizontal pair nn to $i$. In the Ha1 model $i$ must be empty for the adsorption to occur. In the Ha2 model $i$ can be empty, filled with an O, or filled with a CO.}
\label{sk2}
\end{figure}

\begin{figure}[ht]
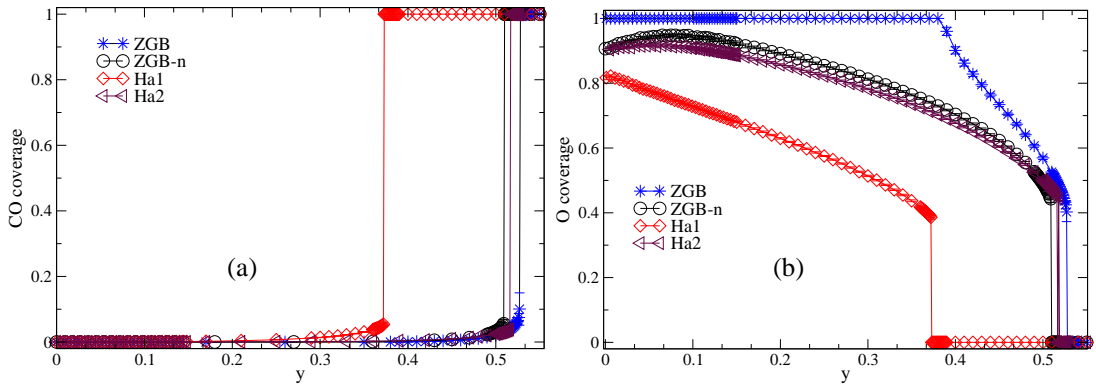

\includegraphics[clip,angle=0,width=.40 \textwidth]{co.eps}
 \includegraphics[clip,angle=0,width=.40 \textwidth]{cox.eps}
\caption{Average values of the CO coverage (a), and the O coverage (b), shown as functions of the stationary applied CO pressure $y$, calculated for the different models.}
\label{fig1}
\end{figure}

\begin{figure}[ht]
\includegraphics[clip,width=16pc]{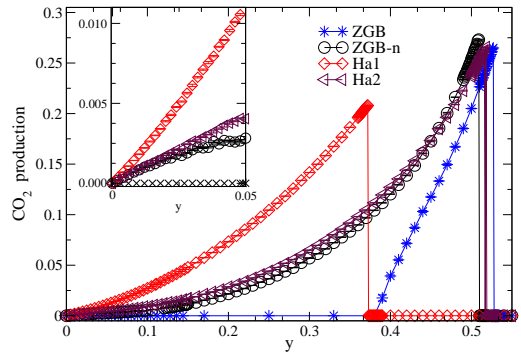}
\caption{CO$_2$ production rate in terms of $y$ for the different models. Notice that for the standard ZGB model the production rate is zero until $y$ reaches a minimum value $y_1\approx 0.38$, while for the other models the production starts as soon as $y >0$ as can be seen in the inset. }
\label{fig2}
\end{figure}

\begin{figure}[ht]
\includegraphics[clip,angle=0,width=.41 \textwidth]{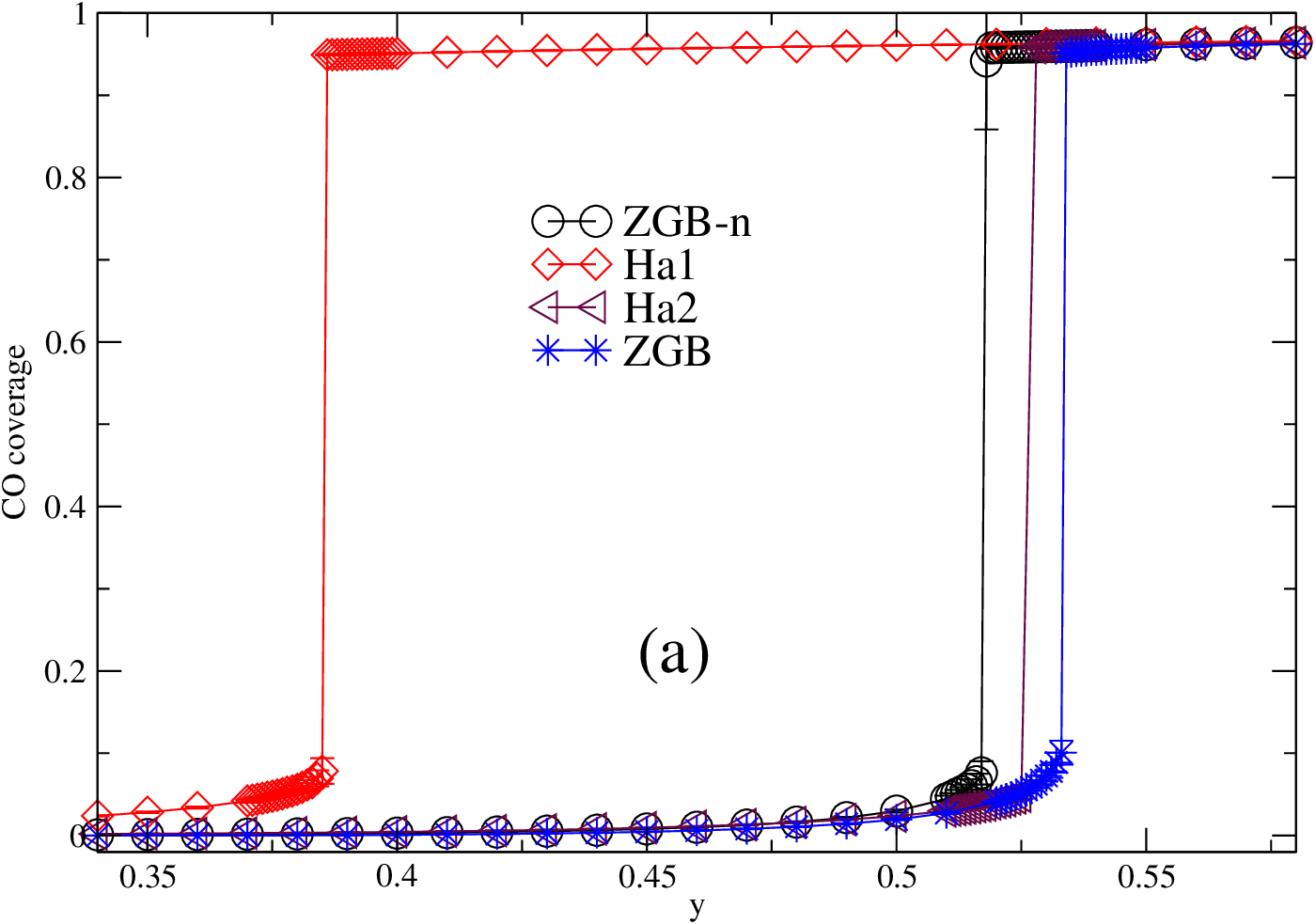}
 \includegraphics[clip,angle=0,width=.41 \textwidth]{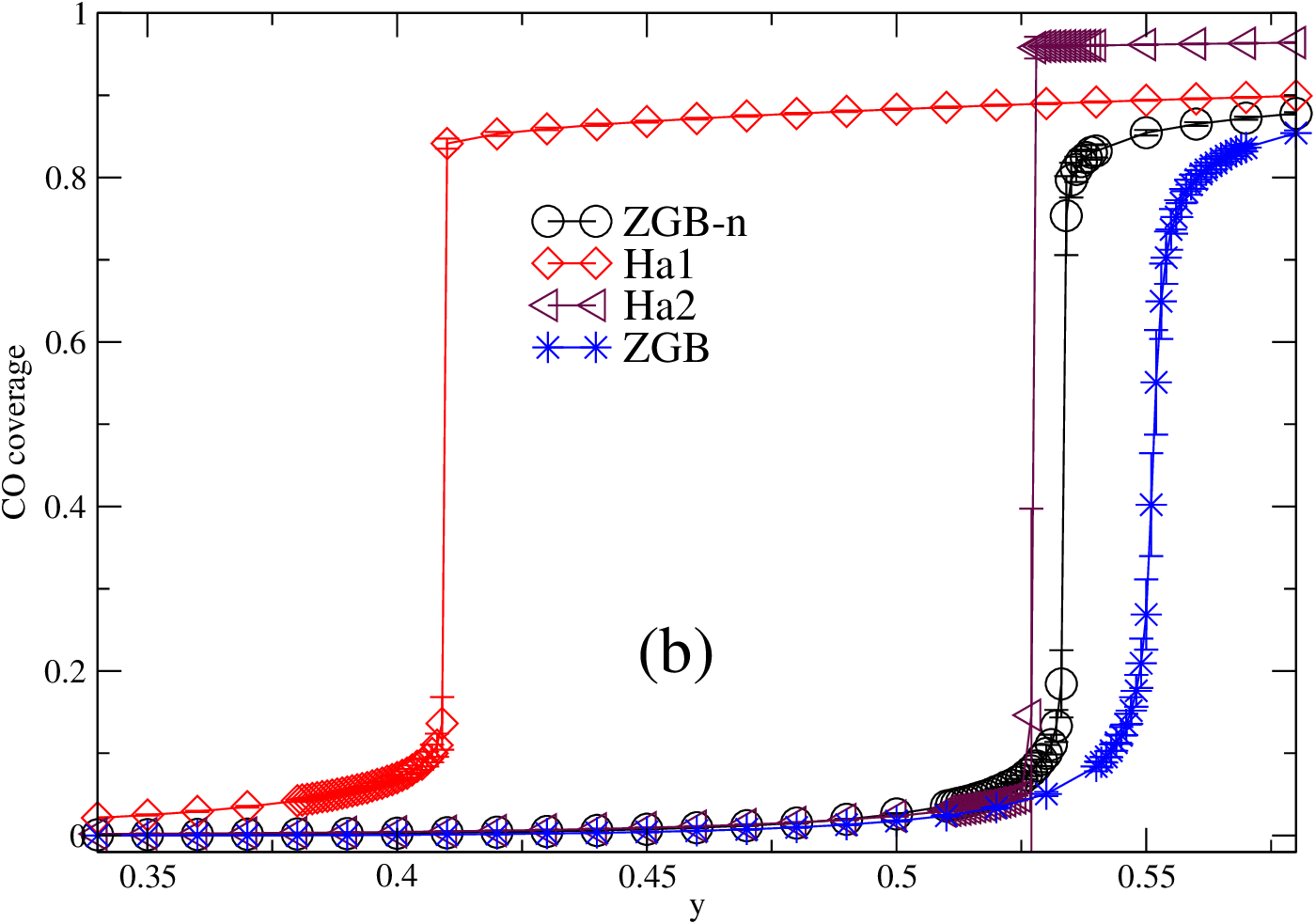}
\caption{CO coverage when there is a CO desorption probability $k$. (a) $k=0.02$ (b) $k=0.06$. }
\label{fig3}
\end{figure}

In this article we have investigated by kinetic Monte Carlo simulations the catalytic oxidation of CO on a surface. We have modified the entrance mechanism of the oxygen atoms in the original ZGB model, to take into account experimental results that indicate that the oxygens tend to come apart after the adsorption. We explore three different entrance mechanisms. In the so-called ZGB-n model the oxygens enter in nnn sites in the surface (see Fig.~\ref{sk1}). In the Ha1 and Ha2 models they enter in the neighboring sites of a selected site that must be empty for the Ha1 and can be filled for the Ha2 (see Fig.~\ref{sk2}). The ZGB model predicts two phase transitions: a continuous one at low values of the CO pressure, and a discontinuous one at higher values of $y$. The continuous transition has not been observed experimentally. We found that for all the proposed mechanisms the nonphysical transition disappears and the continuous transition is slightly shifted toward lower values of $y$, the last effect is particularly evident for the Ha1 model. Finally, in order to make the models more realistic, we include a CO desorption term that reproduces the effect of the temperature. We found that in all the cases the discontinuous transition is shifted toward higher values of $y$, and that the production rate increases with $k$. We think that it is possible that the adsorption mechanisms presented here occur in different real catalytic processes, and we hope that this study can be useful for identifying them.

\vspace{16mm}

\vspace{14mm}

\begin{figure}[ht]
\includegraphics[clip,angle=0,width=.41 \textwidth]{ox_k02.eps}
 \includegraphics[clip,angle=0,width=.41 \textwidth]{ox_k06.eps}
\caption{O coverage when there is a CO desorption probability $k$. (a) $k=0.02$ (b) $k=0.06$.  }
\label{fig4}
\end{figure}

\begin{figure}[ht]
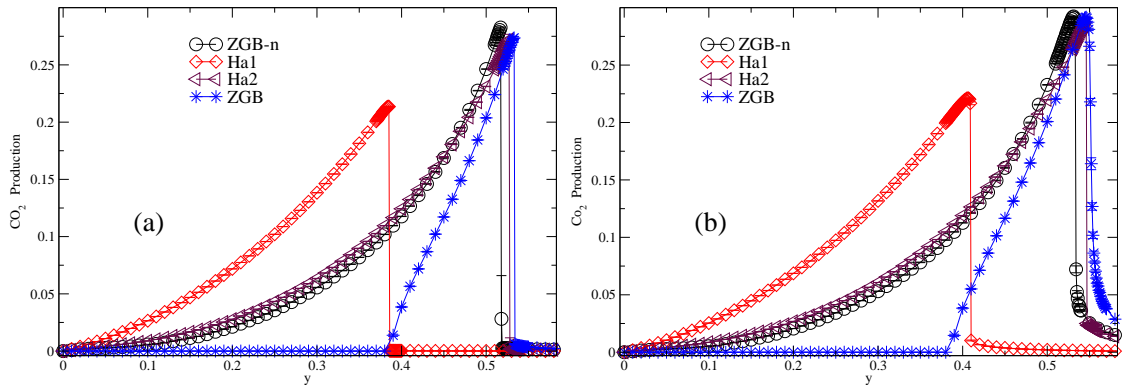

\includegraphics[clip,angle=0,width=.41 \textwidth]{co2k02.eps}
 \includegraphics[clip,angle=0,width=.41 \textwidth]{co2_k06.eps}
\caption{CO$_2$ production when there is a CO desorption probability $k$. (a) $k=0.02$ (b) $k=0.06$.}
\label{fig5}
\end{figure}

\begin{figure}[ht]
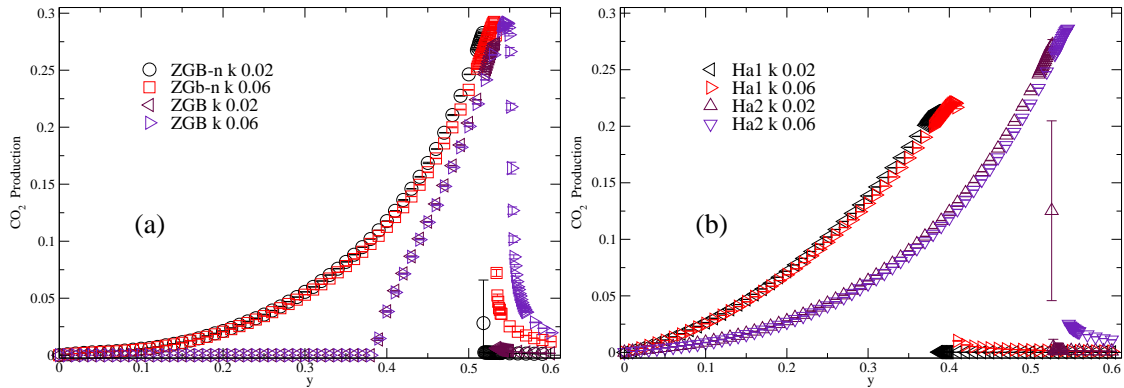

\includegraphics[clip,angle=0,width=.41 \textwidth]{cozcompk.eps}
 \includegraphics[clip,angle=0,width=.41 \textwidth]{co2compk.eps}
\caption{Comparison between the production rates for the different mechanisms for two values of $k$. (a) For the standard ZGB and the ZGB-n model (b) For the Ha1 and the Ha2 models. }
\label{fig6}
\end{figure}

\end{document}